\documentclass[useAMS,usenatbib]{mn2e}
\usepackage{natbib}
\usepackage{longtable}
\usepackage{rotating}
\usepackage{graphics}
\usepackage{array}
\usepackage{graphicx}
\usepackage{amssymb}
\usepackage{array}
\def\mnras{MNRAS}%
\def\aaps{A\&AS}%
\def\aap{A\&A}%

\def\pasj{PASJ}%
\def\apj{ApJ}
\def\aj{AJ}
\def\apjl{ApJ}
\def\iaucirc{IAU~Circ.}%
\def\jqsrt{J.~Quant.~Spec.~Radiat.~Transf.}

\begin{document}


\title[Optical photo-polarimetry of comet 17P/Holmes]{Optical
polarimetry and photometry of comet 17P/Holmes}

\author[U. C. Joshi, S. Ganesh, K. S. Baliyan]{U.C. Joshi\thanks{e-mail: joshi@prl.res.in},
S. Ganesh, and K. S. Baliyan\\
Astronomy \& Astrophysics Division, Physical Research Laboratory,  Navrangpura, Ahmedabad 380009, India}

\date{Submitted to MNRAS }
\pagerange{\pageref{firstpage}--\pageref{lastpage}} \pubyear{2009}
\maketitle
\label{firstpage}

\begin{abstract}

Comet 17P/Holmes was observed for linear polarisation using the
optical polarimeter mounted on the 1.2m telescope atop Gurushikhar
peak near Mt. Abu during the period November-December 2007.
Observations  were  conducted through the IHW narrow band
(continuum) filters.  During the observing run the phase angle was
near $13^{\circ}$  at which the comet showed negative polarisation.
On the basis of the observed polarisation data we find comet
17P/Holmes to be a typical comet with usual dust characteristics. We
note that radial rate of change of brightness in coma in red band is
higher than that in blue band; it has decreased by a factor of 3.6
and 2.5 respectively in red and blue bands during the November -
December run, indicating relative increase in the abundance of
smaller dust particles out ward. Radial brightness variation seen
near the nucleus on November 6 is indicative of the presence of a
blob or shocked region beyond 10" from the nucleus which has
gradually smoothened by December 13. The brightness distribution is
found steeper during November 5-7 as compared to on December
13.

\end{abstract}



\section {Introduction}

Physical properties of cometary dust can be obtained from the
solar radiation scattered by the cometary dust which, in the process, 
gets polarised. The degree of polarisation and its direction mainly
depend on the size distribution, composition of the particles,
phase angle and the wavelength of the incident solar
radiation. However, the real situation is not that straight
forward. In an attempt to study the detailed behaviour of
polarisation with phase, \citet{Dollfus1988} synthesized the 
polarisation observation data on 1P/Halley by various researchers
and derived curves of polarisation as a function of the phase angle.
Phase curve below about $ 20^{\circ}$ shows negative polarisation
while it is positive at higher phase angles. They
find slight modification in the polarisation phase curve as one
moves away from the nucleus indicating the change in the nature of
the dust particles. Also an anomalously high transient polarisation
was noted between October 17 and 30, 1985, at phase angle
$25^{\circ}$, attributed to sudden release of large number
of smaller particles. On the basis of this work on 1P/Halley,
\citet{Dollfus1989} derived the physical properties of the dust grains
indicating the presence of large particles - aggregates comprising
of submicron sized grains, very rough and dark.
These rather large grains are mixed with the clouds of small
particles and they are usually responsible for almost all the
polarisation effects in visible light, except during temporary
specific dust release events (as seen in case of 1P/Halley during
October 17-30, 1985) by the nucleus \citep{Dollfus1989}. The complex
behaviour of the dust is also seen in comet C/1995 (Hale-Bopp), 
especially the region around the nucleus shows complex
structure \citep{hadamcik2003}. Though the comets,
in general, show similar polarisation behaviour with phase angle,
they are divided into three classes based on the maximum in
polarisation\citep{levasseur1999}. Varying polarisation
observed in the coma or in the features (jets, shells, etc.) indicates a
diversity of dust particles. Issues related to the dust
characteristics are adequately reviewed by \citet{kolokolova2005}.\\

One of the main objective behind the study of comets is to
understand the origin of the solar system. Since comets spend
substantial part of their life away from the sun, their sub-surface
material is considered pristine. Space mission Deep Impact was
launched on January 12, 2005 to study the composition of the
interior of the comet 9P/Tempel 1 by colliding a part of the
spacecraft with the comet \citep{hearn2005b, meech2000}. At 5:52
UTC on  July 4, 2005, the impactor of the Deep Impact probe
successfully collided with the comet's nucleus, excavating huge
amount of debris
from the interior of the nucleus \citep{hearn2005a, hearn2005b}. 

On October 24, 2007 nature itself provided a similar
opportunity, albeit at a much grandeur scale. Comet
17P/Holmes underwent unprecedented outburst on 24 October 2007, 
 after about 5 months of perihelion  passage
\citep{santana2007}. The comet brightenings are well documented in
the literature and in general are associated with an
evident fragmentation of cometary nucleus.  
\citep[eg.][and the references therein]{sekanina2002a,
sekanina2002b}. But the magnitude and the characteristics of the
outburst occurred in comet 17P/Holmes has dwarfed all the events
seen earlier.

 The periodic comet 17P/Holmes was discovered on November 6, 1892
by E. Holmes during a huge outburst  which was followed by another
similar event in January 1893. When discovered, the comet was
around five months past perihelion. The October 24, 2007 outburst
also occurred about six months after the perihelion passage (May 4,
2007) of the comet. The heliocentric distances of the comet
at the time of the two outbursts were nearly similar (2.39 and 2.44
AU respectively in 1892 and 2007).
 During the outburst huge amount of dust and gas 
ejected out from the comet's sub-surface. Comet 17P has typical
composition as discussed by \citet{hearn1995} but  the dust-to-gas
ratio has been found very high on November 1, 2007
\citep{schleicher2007} and is attributed either due to the
finite lifetimes of the gas molecules released at the onset of the
outburst as compared to the long lived dust grains or due to a
portion of the dust tail remaining within the photo-center apertures
due to projection effects from the comet's small phase angle.  

As discussed above, polarisation studies have proven to be
very useful in inferring the dust properties of comets.  While the
data at large phase angles are substantive, there is a lack of the
observations at small phase angles which hinders study of the
polarisation behaviour in the negative branch of polarisation-phase
curve. In the past we have made detailed study on comets  Hale-Bopp
and WM1/LINEAR at low phase angles \citep{Ganesh1998,Joshi2002,
Joshi2003}. In the present communication we present results on the
polarisation and photometric observations and discuss the brightness
distribution across the coma and its evolution in the comet
17P/Holmes during the period November 5-7 and December 13, 2007
when the phase angle was near $13^{\circ}$. We compare our
observations with recently reported results on this comet by
\citet{rosenbush2009} and also discuss polarisation behaviour with that of
other comets observed at similar phase angles.

\begin{figure*} 
\centering{
\includegraphics[width=\textwidth]{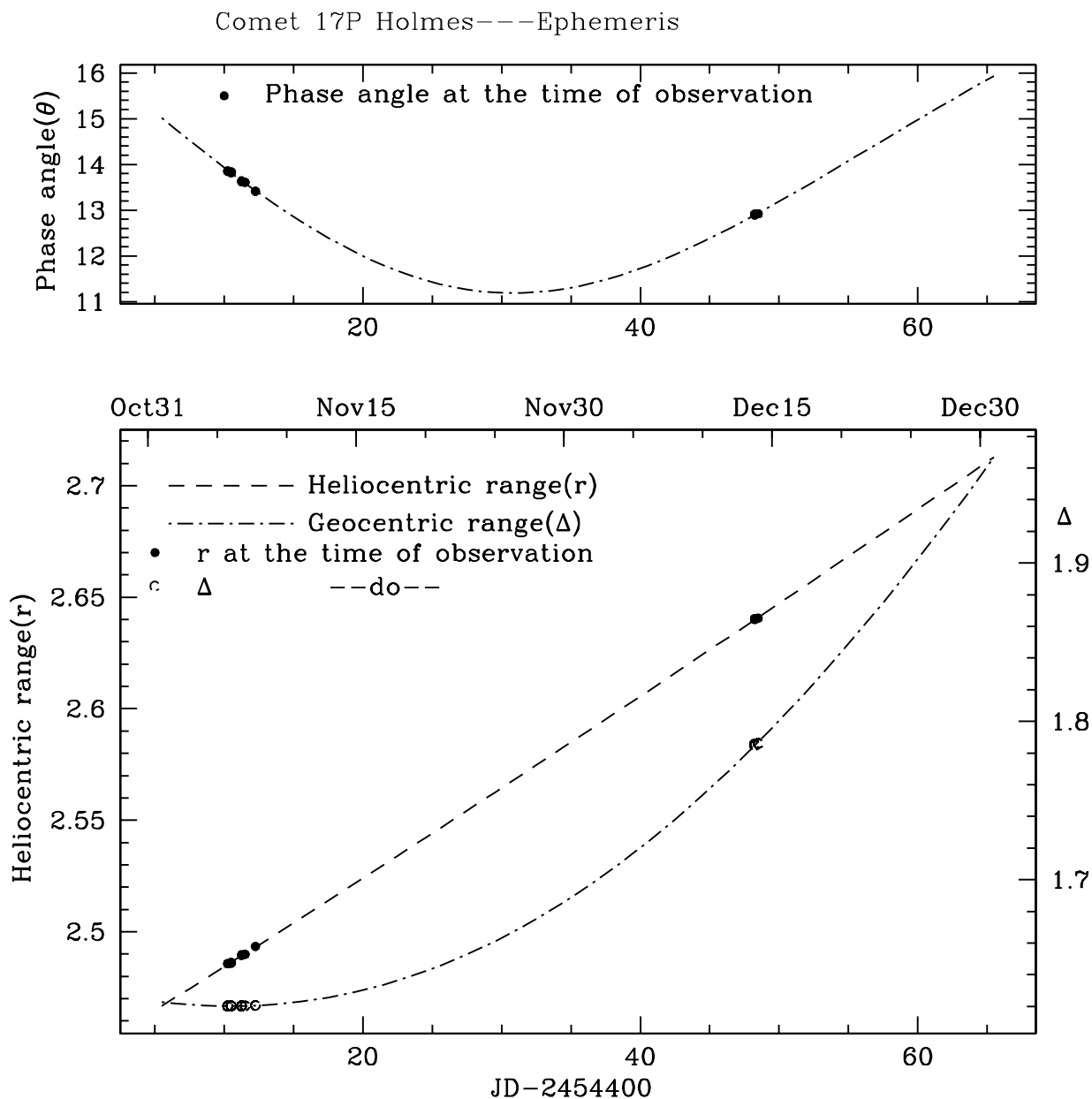}}
\caption{Top panel shows phase angle ($\alpha$) while bottom panel shows the apparent heliocentric (r), and geocentric ($\Delta$) ranges relative to the observer during the observing run.}
\label{17P_ephe}
\end{figure*}
\section{Observations and analysis}

Photopolarimetric observations of comet 17P/Holmes were made during the period
November 5-7 and on December 13, 2007 with a two channel photo polarimeter
\citep{Deshpande1985,Joshi1987}, which has been fully automated recently
\citep{ganesh2008p}, mounted on the 1.2m telescope of Mt. Abu Observatory
operated by Physical Research Laboratory (PRL), Ahmedabad. {\bf PRL polarimeter
works on rapid modulation principle. A rotating  super-achromatic half-wave
plate in front of a fixed Wollaston prism modulates the polarized component of
the incident light at a frequency of 10Hz. Another identical Panchratnam
half-wave plate is put between the rotating half-wave plate and the Wollaston
prism to eliminate any wavelength dependence of optic axis of half-wave plate.
Rotating half plate rotates at discrete steps with 1 ms sampling time. One full
rotation of half-wave plate is completed in 96 step and thus one modulation
cycle involves 24 steps (i.e. 24ms) and the data are folded and accumulated in
24 bins. With the rapid modulation, the atmospheric scintillation effects (eg.
sky transparency fluctuations) are eliminated. \\}

 The instrument is equipped with IAU's International Halley
Watch ($IHW$)) continuum filters (3650/80\AA, 4845/65\AA, 6840/90\AA~)
\citep{Osborn1990} and Johnson-Cousins' BVR broad band filters. The
$IHW$ filters acquired for Comet Halley have been used for these
observations.  These filters have been in regular use for observations
of several other comets \citep{Joshi1987,Sen1991,Ganesh1998,Joshi2002,
Joshi2003}. The filters have been carefully stored in dry atmospheric
conditions to preserve their transmission characteristics. The
observations made with the same set of filters facilitate better
comparison with other comets observed  earlier,  hence their continued
use is justified.\\ 

{\bf The online data reduction performed after each integration invokes a
least square fit to the counts (comet-sky) obtained from the two
photo-multiplier tubes as a function of the rotating half-wave plate
position. The mean error in polarisation is estimated from actual
deviation of the counts from the fitted curve. To further reduce the
error, several such measurements were taken and averaged. The error
bars represent this error and instrumental error (obtained by observing
several zero polarisation standards during each night and  found to be
$<0.03\%$).  Regarding the sky polarisation, on an average it was $\sim
5-6\% (\pm 3\%)$ but the sky as seen through
the 26" aperture is $\sim 4$ mag fainter compared to the comet 
observed through the same aperture. These numbers were 
consistent during the observing run. Hence, the error communicated due
to the sky to the comet data is negligible.} Nonetheless, to take care
of the sky polarisation, observations were made alternately centred on
the photo Centre of the comet and on a region of the sky more than 30
arcmin away from the comet (along the anti-tail direction).  All the
observations were made under dark sky conditions.
The errors in the position angle are obtained using the equation 8.5.4
given by \citet{Serkowski1974}.\\ 

The observations were taken with apertures (non-metallic diaphragms
centred on the photo-centre of the comet) of different sizes- 10", 20",
26" and 54" with the projected diameter varying from 11750 to 69925 km (see Table
\ref{obstab})  to study the behaviour of the dust as a function of radial
distance from the comet nucleus. All the observations were made
under dark sky conditions, and the  comet was much brighter than the sky
resulting in negligible contribution by the sky to the
observed degree of polarisation of the comet. Nonetheless, to take care
of the sky polarisation, observations were made alternately centred on
the photo Centre of the comet and on a region of the sky more than 30
arcmin away from the comet  (along the anti-tail direction).\\  

Polarisation standard 9 Gem was observed to calibrate the observed
position angle. Comet's IHW magnitudes were obtained using the
observations of solar type stars, namely HD29461, HD76151.  Polarisation values, corrected position angle and
IHW magnitudes in continuum bands are given in Table \ref{obstab}. 

\begin{table*} 
\caption{Polarisation observations of comet 17P/Holmes. Listed
 entries are Julian date(JDT), Heliocentric range(r), Geocentric range($\Delta$),
 phase angle($\alpha$), filter, aperture(arcsec), Diameter(km),
 total integration time(IT sec), degree of polarisation($P\%$), error in 
 polarisation(${\epsilon}_P\%$), position angle($\theta^\circ$) in equatorial plane,
 magnitude at the time of observations. JDT=JD-2454400}
\begin{tabular}{|l|l|l|l|l|l|l|l|l|l|l|l|l|} 
\hline
Date&JDT& r&${\Delta}$&${\alpha}$& Filter(\AA)& Ap(")&Dia(km)& IT(sec)& $P\%$ & ${\epsilon}_p\%$ &${\theta^\circ}$ & mag \\
\hline
Nov 5&10.32687 & 2.485602 & 1.620252  & 13.9 & 6840 &  10 & 11751 & 400 &  -1.17 &  0.50 & 19 & 10.05\\ 
&10.33699 & 2.485642 & 1.620252  & 13.9 & 4845 &  10 & 11751 & 300 &  -1.70 &  0.51 &  4 & 11.09\\ 
&10.34764 & 2.485684 & 1.620251  & 13.9 & 3650 &  10 & 11751 & 500 &  -2.72 &  2.14 & 81 & 12.45\\ 
&10.35990 & 2.485732 & 1.620251  & 13.8 & 6840 &  26 & 30553 & 400 &  -1.22 &  0.21 & 24 &  8.24\\ 
&10.36958 & 2.485770 & 1.620252  & 13.8 & 4845 &  26 & 30553 & 400 &  -1.09 &  0.18 & 22 &  9.28\\ 
&10.38014 & 2.485791 & 1.620252  & 13.8 & 3650 &  20 & 23502 & 100 &  -1.25 &  0.76 & 86 & 10.58\\ 
Nov 6&11.28874 & 2.489386 & 1.620319  & 13.6 & 6840 &  26 & 30554 & 500 &  -1.60 &  0.27 & 14 &  8.66\\ 
     &11.30002 & 2.489430 & 1.620321  & 13.6 & 4845 &  26 & 30554 & 500 &  -0.95 &  0.19 & 20 &  9.72\\ 
     &11.31650 & 2.489495 & 1.620323  & 13.6 & 3650 &  26 & 30554 & 900 &  -1.51 &  0.70 & 32 & 10.98\\ 
     &11.33392 & 2.489564 & 1.620327  & 13.6 & 3650 &  10 & 11751 & 400 &  -2.60 &  2.15 & 15 & 12.56\\ 
     &11.34823 & 2.489620 & 1.620330  & 13.6 & 4845 &  10 & 11751 & 800 &  -0.96 &  0.43 & 22 & 11.25\\ 
     &11.36806 & 2.489698 & 1.620334  & 13.6 & 6840 &  10 & 11751 & 700 &  -1.10 &  0.54 & 43 & 10.16\\ 
     &11.38504 & 2.489765 & 1.620338  & 13.6 & 6840 &  20 & 23503 & 600 &  -1.49 &  0.29 & 20 &  9.01\\ 
     &11.40166 & 2.489831 & 1.620343  & 13.6 & 4845 &  20 & 23503 & 400 &  -1.36 &  0.27 & 20 & 10.05\\ 
     &11.41631 & 2.489888 & 1.620348  & 13.6 & 6840 &  54 & 63460 & 400 &  -1.09 &  0.14 & 13 &  7.21\\ 
Nov 7&12.30185 & 2.493380 & 1.620604  & 13.4 & 6840 &  26 & 30559 & 600 &  -1.60 &  0.23 & 15 &  8.67\\ 
     &12.32505 & 2.493472 & 1.620613  & 13.4 & 4845 &  26 & 30560 & 600 &  -1.17 &  0.16 & 17 &  9.73\\ 
     &12.33843 & 2.493524 & 1.620618  & 13.4 & 3650 &  26 & 30560 & 500 &  -1.02 &  0.71 & 97 & 11.03\\ 
Dec 13&48.25186 & 2.639940 & 1.784706  & 12.9 & 6840 &  10 & 12944 & 600 &  -3.96 &  1.77 & 97 & 11.69\\ 
     &48.26690 & 2.640003 & 1.784838  & 12.9 & 4845 &  10 & 12944 & 600 &  -1.39 &  0.76 & 96 & 12.16\\ 
     &48.28473 & 2.640077 & 1.784996  & 12.9 & 3650 &  10 & 12946 & 600 &  -2.88 &  3.46 &142 & 12.98\\ 
     &48.29981 & 2.640141 & 1.785130  & 12.9 & 3650 &  26 & 33662 & 300 &  -3.20 &  4.12 &131 & 12.72\\ 
     &48.31233 & 2.640193 & 1.785241  & 12.9 & 3650 &  20 & 25895 & 300 &  -3.52 &  3.55 &138 & 12.84\\ 
     &48.32681 & 2.640254 & 1.785370  & 12.9 & 3650 &  54 & 69923 & 400 &  -4.56 &  3.53 &107 & 12.49\\ 
     &48.33570 & 2.640291 & 1.785449  & 12.9 & 4845 &  54 & 69926 & 300 &  -1.79 &  0.88 &114 & 11.55\\ 
     &48.34532 & 2.640331 & 1.785535  & 12.9 & 4845 &  26 & 33669 & 400 &  -1.97 &  0.83 &110 & 11.92\\ 
     &48.35515 & 2.640372 & 1.785623  & 12.9 & 4845 &  20 & 25901 & 400 &  -1.60 &  0.79 & 64 & 12.02\\ 

\hline 
\end{tabular} 
\label{obstab} 
\end{table*}
\section{Results \& discussion} 
The various quantities obtained from our observations on the comet
17P/Holmes are listed in Table \ref{obstab} namely, the degree of
polarisation($P\%$) and its error($\epsilon_{P}\%$), the position
angle($\theta)$, brightness magnitudes along with observing
time(JDT),
filter, aperture size, total integration time, and the orbital
parameters at the time of  observation. All the values listed in
Table \ref{obstab} have been carefully checked for any inconsistency.
Figure \ref{17P_ephe} shows the Heliocentric and Geocentric ranges
and the phase angle($\alpha$) at the time of observation, which were
obtained using the JPL's HORIZONS system \citep{yeoman}. As mentioned
earlier in section 2, we have made observations through various
apertures for sampling the comet. However, 26" aperture is used more
often and if not mentioned specifically, we will use this aperture
for further discussion. This aperture corresponds to a projected 
diameters $\sim$30550km during November 5-7 and $\sim$33670 km on
December 13, 2008. The sampled area is large and while small scale
inhomogeneities are expected to average out, large structures in coma
might still show up. The polarisation values in the blue narrow band
are associated with larger errors due to poor $S/N$ ratio. To
address the possibility of continuum BC band contamination by
molecular emission, which might affect the degree of polarisation, we
used the spectra provided by Buil (private communication)\footnote
{see his web-site http://www.astrosurf.com/buil/}. Examining the
spectrum of November 1, a very weak emission feature ($<3\%$ of the
continuum) appears to partly overlap with the BC band. Even if we
take the upper limit of $3\%$ contamination of the BC band by this
feature, it will change the degree of polarisation by $<0.05\%$. 
In fact in the present case, any such correction will increase 
the absolute value of the polarisation that will improve the fit. 
Molecular production rate in 17P/Holmes is reported to decay
exponentially with time \citep{schleicher2009}, the contribution of
molecular emission will further reduce at later date (e.g. November
5-7 and December 13) compared to November 1, 2007 when the spectrum
was taken. Hence we have ignored the contamination of the BC band by
molecular emission.\\   

During our observing run, the phase angle remained less than
$20^\circ$ and observed polarisation is negative. 

 In the following we discuss the wavelength and phase angle
dependence of polarisation and compare the present results with 
the results of \citet{rosenbush2009}.

\begin{figure*} 
\includegraphics[width=0.45\textwidth]{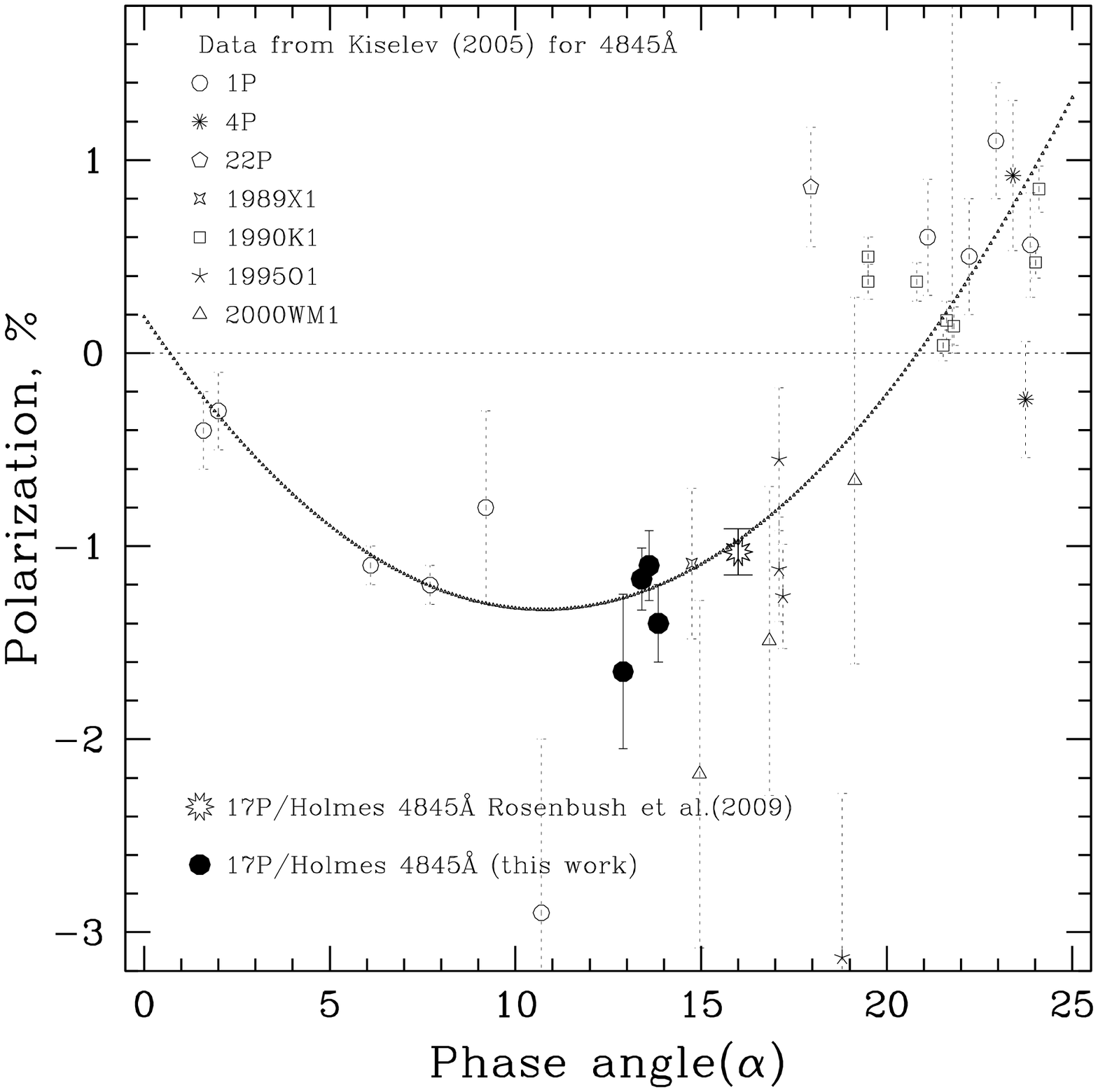}
\includegraphics[width=0.45\textwidth]{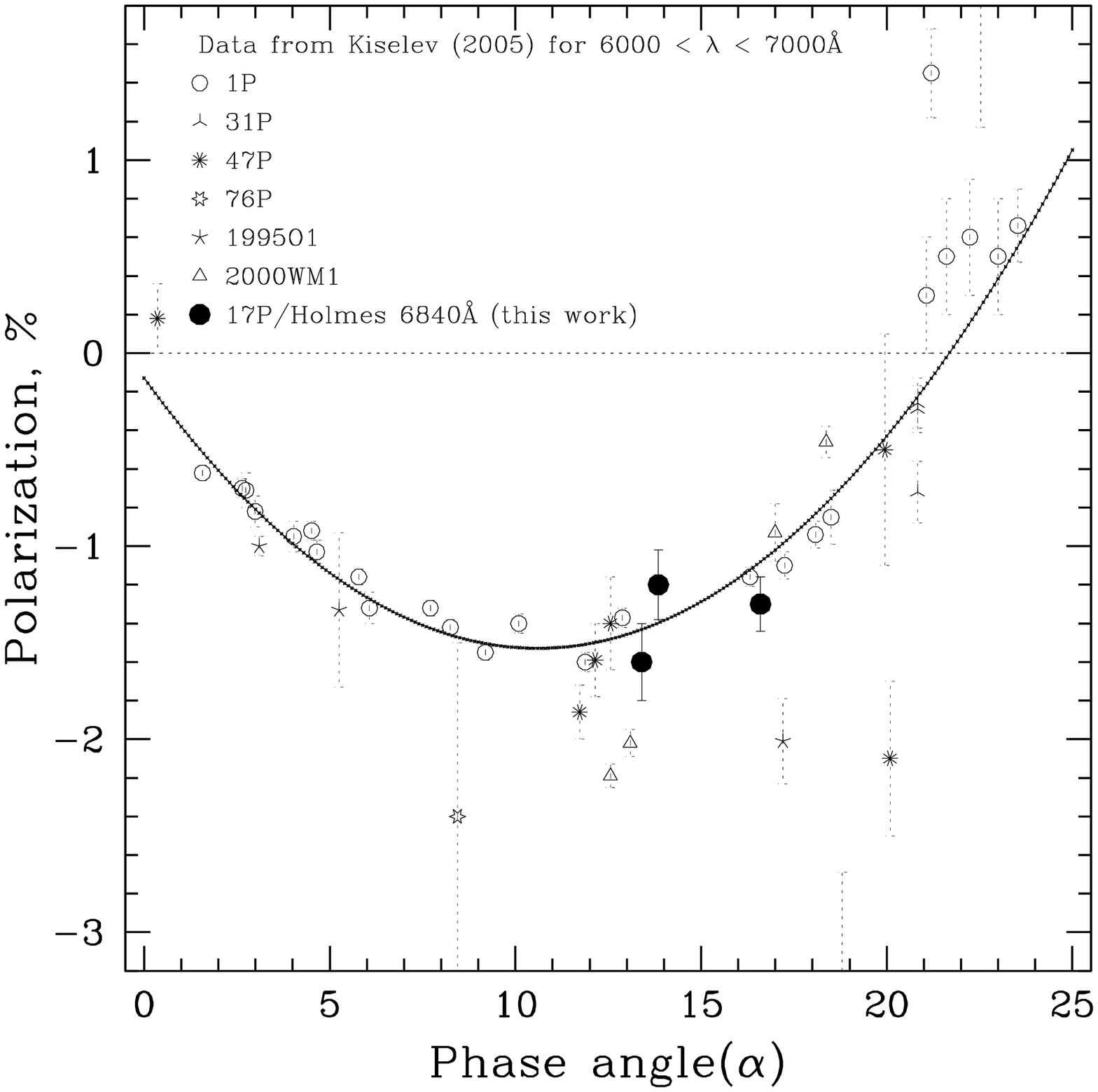}

\caption{Polarisation vs phase plot. (Left panel) $4845\AA$ filter data from
\citet{kiselevCat} and this work; (Right panel) same as in left panel but for
wave length range $6000<\lambda<7000\AA$. Observations are labeled
with different symbols for all comets. 2nd order polynomial fit to
the data from the catalogue is shown as solid curve. In case of
multiple observations in the catalogue at the same phase, an average
value is taken; observations with uncertain position angle
measurements are not included.}
\label{PP6840}
\end{figure*}

\subsection {Polarisation - phase angle dependence} Figure
\ref{PP6840} presents P\% vs $\alpha$ curve for $\alpha <
25^\circ$ for the comets in the blue band (4845\AA) and the red
band ($6000<\lambda<7000\AA$). Since our observations are made
at low phase angles where the polarisation is negative, we  have
limited the plots to phase angle $< 25^{\circ}$. Also plotted are
the data from Kiselev's catalogue \citep{kiselevCat} for the comets
which have been observed by various researchers in blue and
red-bands(all narrow and broad bands). As observations in narrow
continuum bands at the low phase angles are scanty, we considered
all the observations made in red filters ($6000<\lambda<7000\AA$) to
generate an average P\% vs $\alpha$ curve  for $\alpha < 25^\circ$.
Unpublished data, as mentioned in the catalogue, and the data for
which P\%$> 0.0$ for $\alpha < 20^{\circ}$ and the outliers from
the main trend have been ignored. The best polynomial fit thus
generated is presented in Figure~\ref{PP6840} with solid
curve.\\ 

The present observed values of the polarisation in 17P are  shown
in two curves with solid circles with error bars ($\pm \sigma$).
For plotting purpose, all the data obtained through the same filter
but with different apertures are averaged taking the weight equal
to $1/\sigma^2$. In both, red and blue domain, our observed
polarisation values lie close to the best fit curves.  These
observations are suggestive of usual nature of the comet
17P/Holmes.

\begin{figure*}
\centering{
\includegraphics[width=\textwidth]{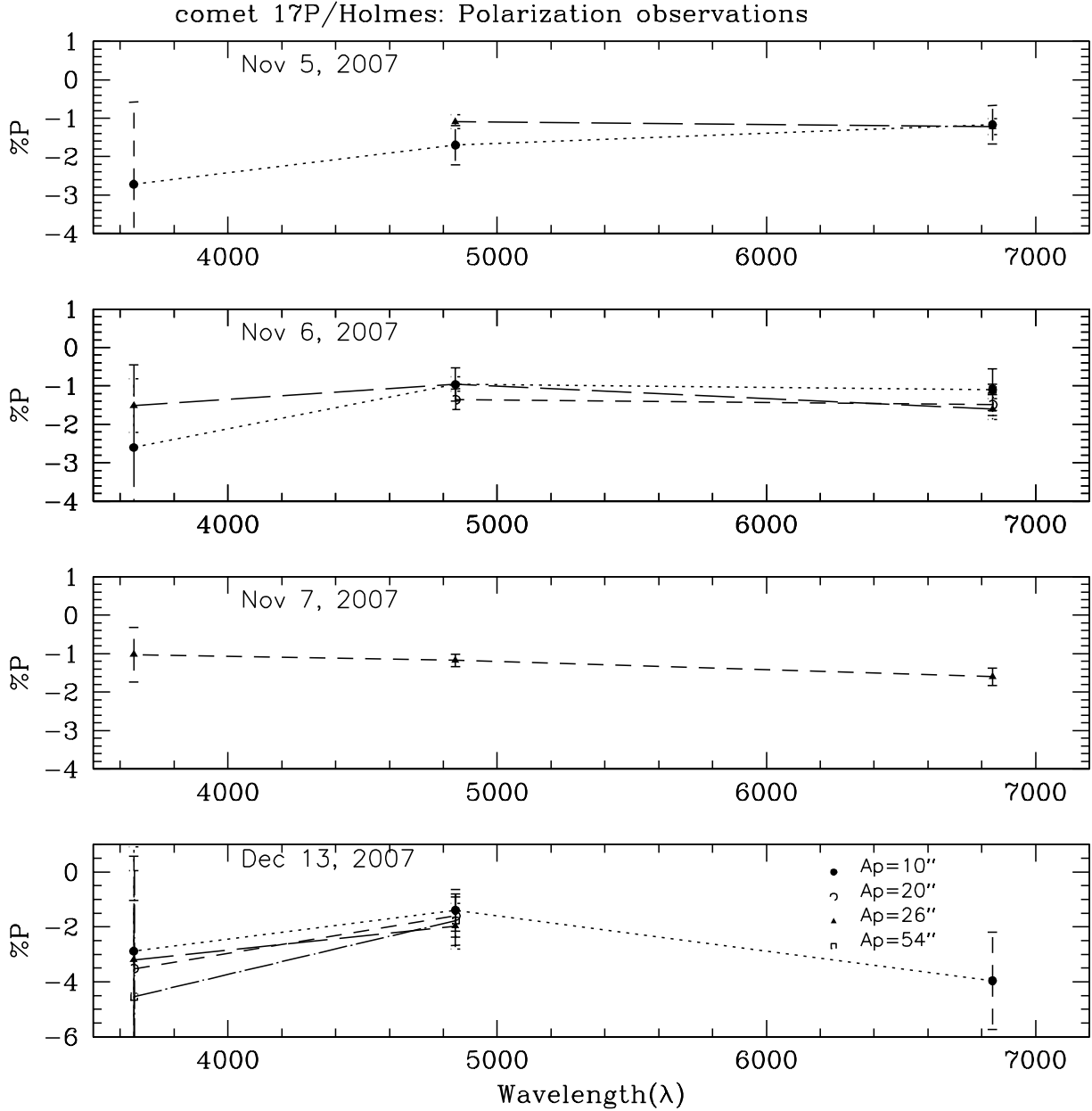}}

\caption{Wavelength dependence of polarisation as observed using the continuum
filters through different apertures. Observing date is marked on each panel 
and the bottom panel shows annotation used for different apertures. 
Observed points are joined with different type of lines: dot for 10", 
short dash for 20", long dash 
for 26" and dot-long dash for 54" apertures to distinguish them.}
\label{wdep}
\end{figure*}

\subsection{Wavelength dependence of polarisation}
Though the negative branch of polarisation-phase($P\% - \alpha$)
curve is not very sensitive to the change in wavelength($\lambda$), the
composition might lead to a  mild dependence. To study any such
wavelength dependence of polarisation, observed $P\%$ in continuum
bands is plotted against the wavelength in Figure
\ref{wdep}. The observation date, the phase angle and the aperture
used are marked in the figure. The observations made though
different apertures are plotted with different symbols. As expected
for the low phase angle, polarisation is negative(ie the
polarisation vector lies in the scattering plane). Looking at
Figure \ref{wdep}, we do not see, within the observed errors, any
significant dependence of $P\%$ either on wavelength or on the
aperture size. However, one can notice a mild trend of increase in
$P\%$  (decrease in the absolute value of polarisation) on November 5
in the observation with 10" aperture and a mild decrease in $P\%$ on
November 7 trough 20" aperture. This might indicate time evolution
of the cometary grains. As discussed above the $\lambda$-dependence of
P\% is to be taken cautiously.\\

 On October 25, the polarisation is found to change 
 from $-0.7\%$ in 4845-band to $0.0\%$ in
6840-band while on  October 27 and November 3, the polarisation 
does not show any change with wavelength \citep{zubko2009}; its
value remaining at $-1.0\%$ level. This has been attributed to the decrease
in the number density of smaller particles. We also do not notice any
significant change in polarisation with wavelength during 
Nov 3-7 and on Dec 13, 2007. However, \citet{rosenbush2009}
 have reported steep wavelength dependence of polarisation 
 during the period October 27-November
22, 2007. The possible reason for this discrepancy is 
further discussed in section 3.3.\\ 

\subsection{Comparison with other results}
\citet{rosenbush2009} have reported the results based on the
polarisation observations made during the period October 27-November
22, 2007. In the following we compare our observations with theirs
and comment on the findings.
\begin{enumerate}
\item {On November 5, 2007(phase angle $\sim 13.8^\circ$),
\citet{rosenbush2009} have reported P\% in WRC filter while we observed
in BC and RC bands. It is noticed that absolute value of degree of
the linear polarisation reported by them are systematically lower than
our values.}\\
\item {Using their observations in WRC(7228/1140\AA) and
BC(4845/65\AA) bands on October 27, 2007, \citet{rosenbush2009}
estimate large (-0.77/1000\AA) spectral gradient. Though the
spectral trend in the negative polarisation regime is not clearly
established, they term it as atypical spectral behaviour not shown
by normal comets. Throughout our observations in RC and BC bands,
we do not notice such 'atypical' trend.}

\end{enumerate}

Based on their observations of relatively much lower absolute
polarisation and atypical spectral behaviour, they claim comet
17P/Holmes 
to be of unusual type which is contrary to our findings. In the
light of this, we very carefully checked our observations for any
systematic error, sky subtraction etc both for the comet and
polarisation standard star observations. We find our results to
be thoroughly consistent. The average degree of polarisation
during November 5-7 is $\sim -1.2\%$ but on December 13 $P\%$
appears to go  below -1.5\%. These values are close to the values
generally observed in other comets at such  low phase angles, eg.
comet Halley \citep{Sen1991},  Hale-Bopp \citep {Ganesh1998},
LINEAR WM1 \citep{Joshi2002,Joshi2003}. 
The data points lie close to the
best fit (P\% vs $\alpha$ curve), to further supporting the
argument that 17P/Holmes is a usual comet. On the other hand,
\cite{rosenbush2009} have made almost all their observations in
broad bands(WRC, R, I). Their only observation in BC 'narrow'
band on October 27, 2007 gives polarisation value which is very
close (within $1\sigma$) to the typical phase-polarisation curve
for comets(cf Figure \ref{PP6840}) and also agree with
polarisation values obtained by us at later date. It is, therefore,
very likely that the broad band continuum flux  is contaminated
by the gas emission, lowering the absolute value of the degree of
polarisation. Strong gas emission has been detected during the
outburst of 17P/Holmes \citep[e.g.][]{bockelee2008a,
bockelee2008apj, biver2008, schleicher2007}, which supports the
argument given above for the lower absolute value of the
polarisation reported by \citet{rosenbush2009}. \\

Though the comet 17P/Holmes has high dust-to-gas ratio, its abundance
has 'typical' composition \citep{schleicher2007}. Therefore,
within the limitations of the observed polarisation data, it
appears that 17P/Holmes is similar to other typical comets and
follows the average P\% - $\alpha$ curve at low phase angles (ie
$\alpha < 20$). The gas production rates and the dust
characteristics in 17P/Holmes studied by others using narrow band photometry
and spectroscopy indicate similarity between 17P/Holmes and other
typical comets \citep[e.g.][]{schleicher2009, biver2008,
lin2009, watanabe2009}. These further support the inference drawn in the
present study on the basis of polarisation data.

\subsection{Spectral Energy Distribution and the light curve}
\begin{figure*} 
\centering{
\includegraphics[width=\textwidth]{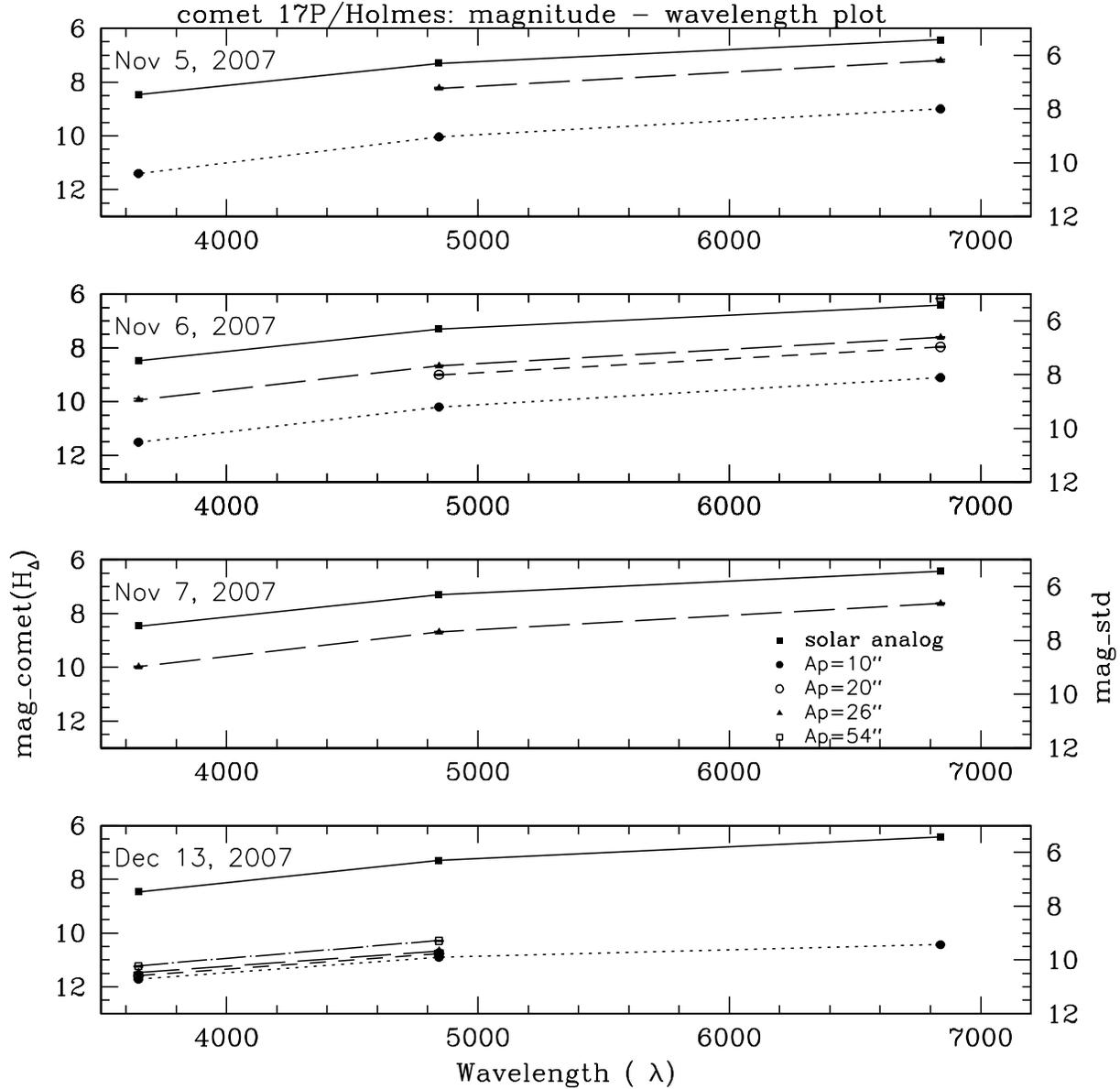}}
\caption{Spectral energy distribution as observed through 
different apertures on different dates. $H_{\Delta}$ is the comet
magnitude referred to the geocentric distance($\Delta$) equal to 1AU.
Error bars $\pm 1\sigma$ are plotted, but these being small ($<$ 0.06mag),
lie with in the symbol used in the plot.}
\label{sed}
\end{figure*}

\begin{figure*} 
\centering{
\includegraphics[width=\textwidth]{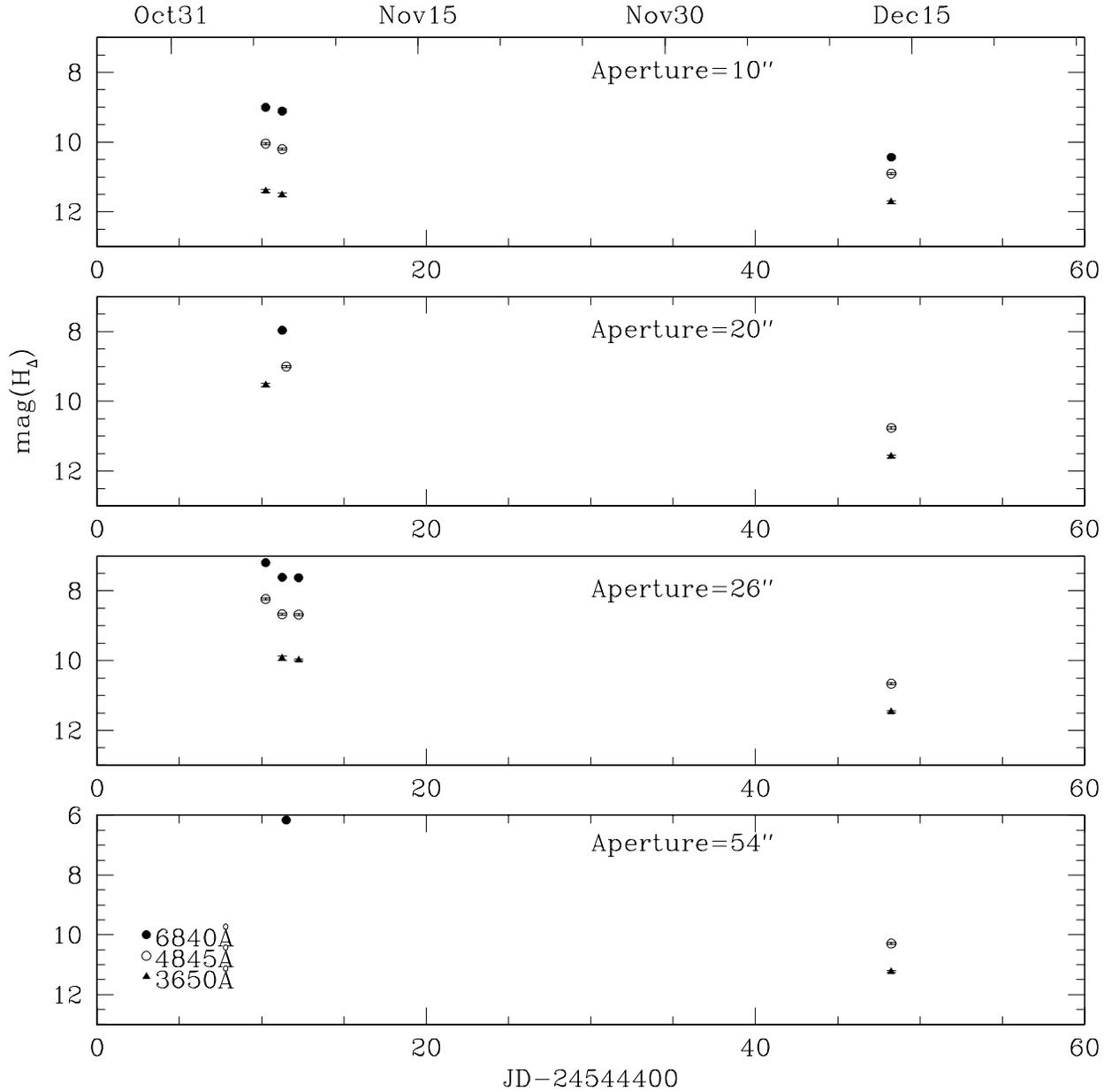}}
\caption{Light curve showing IHW magnitude variation with time(JD). 
 Observed magnitudes
through different size apertures are plotted in different panels.
Error bars ($\pm 1\sigma$), which lie within the symbol used, are plotted.}
\label{light_curve}
\end{figure*}

 \begin{figure*} 
\centering{
\includegraphics[width=\textwidth]{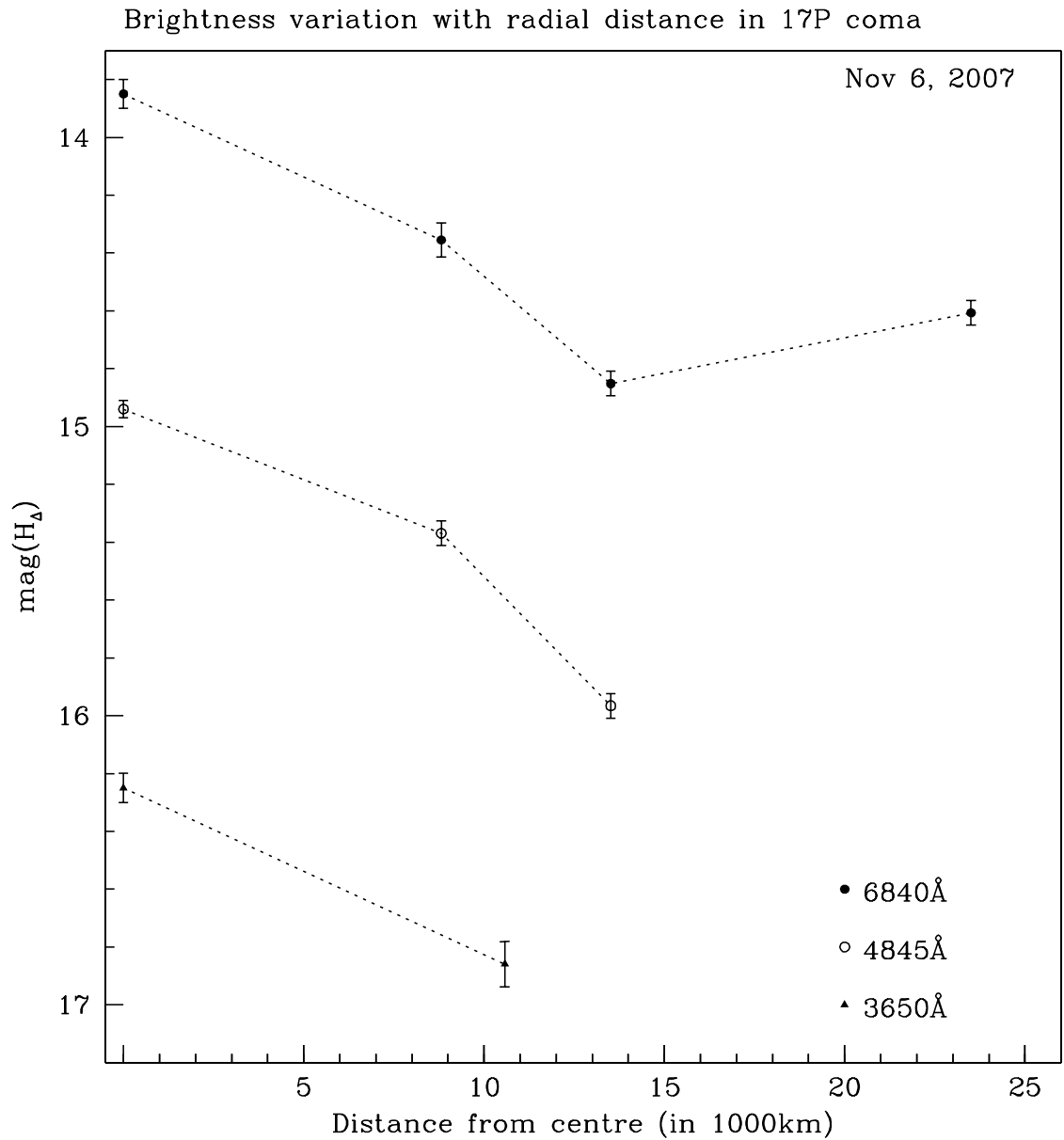}}
\caption{Radial brightness variation as projected on the sky 
on Nov 6, 2007. Error bars are $\pm \sigma$.}
\label{rad_brightness_nov6}
\end{figure*}

\begin{figure*} 
\centering{
\includegraphics[width=\textwidth]{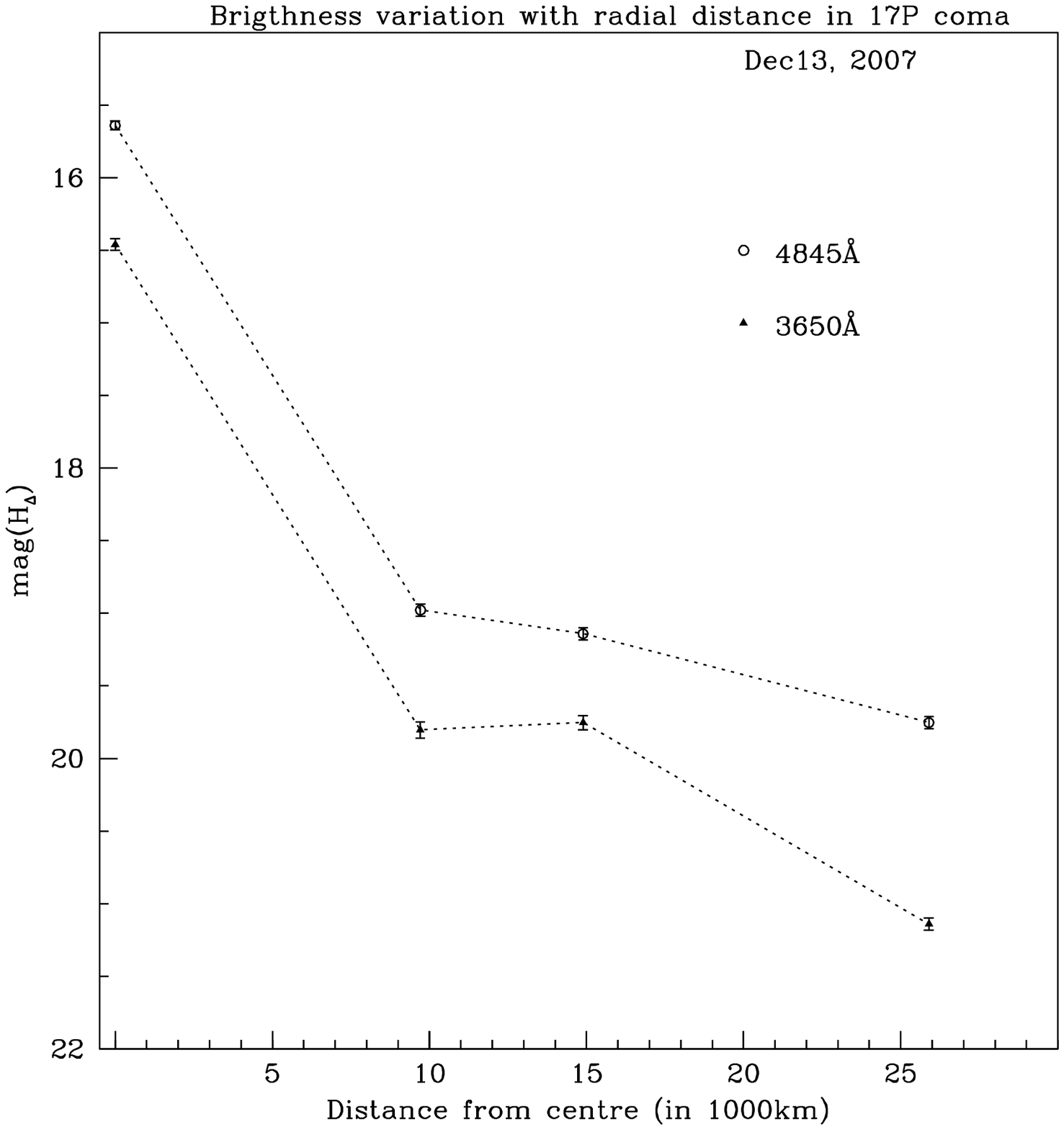}}
\caption{Radial brightness variation as projected on the sky 
on Dec 13, 2007. Error bars are ($\pm \sigma$).}
\label{rad_brightness_dec13}
\end{figure*}

Figure \ref{sed} shows the continuum energy distribution of 17P/Holmes.
In this figure we have plotted normalized magnitudes
$(H_{\Delta}$) which are corrected for extinction and
instrumental effects and referred to a geocentric distance
($\Delta$) of 1AU by subtracting the term 5Log($\Delta$). For
comparison, energy distribution of a solar analog star(HD76151),
which was observed during the observing run, is  also plotted.
 It is seen (cf. Figure \ref{sed}) that 
during the period November 5-7, 2008 the comet colour is similar
to the solar colour. However, on December 13 the colour is
marginally bluer, indicating the increase in relative abundance
of small particles. This can be explained assuming  that in due
course the larger  particles might have fragmented to smaller
particles thus  enhancing the relative abundance of smaller
particles.

Let us now discuss the light curve of comet 17P/Holmes. Figure \ref{light_curve}
shows the evolution of brightness in continuum bands through various size
apertures, plotted in different
panels. Clearly there is decrease in
brightness with time, but the interesting finding is the relative dimming 
in red band being more compared to blue band which is  more apparent in
10" aperture. The brightness of central coma(10") has reduced by $\sim
1.4$ and 1 mag in red and blue wave band amounting to reduction by
a factor of $\sim$ 3.6 and 2.5 in respective bands. The mean spectrum is bluer on December 13, 2007  compared to what
is observed during November 5-7, 2007. 

Observations on November 6 and December 13 are made through four
apertures (10", 20", 26", and 54") which allowed us to investigate the radial distribution
of brightness. A simple approach is adopted to estimate the
brightness at a particular radial distance  from the nucleus (the
brightest point is taken as the nucleus). Brightness in the
annular region is estimated by subtracting brightness in 
successive apertures which is then normalized to one $arcsec^2$ 
and this brightness is assigned to the mean annular distance
between the two apertures. The radial brightness distribution thus
obtained is plotted in Figures  \ref{rad_brightness_nov6} and
\ref{rad_brightness_dec13} respectively. On November 6, the
brightness in the inner region of the coma, as expected, is decreasing
with radial distance, but beyond 12 arcsec from the  nucleus it
starts increasing, indicating the distribution of light to be non
uniform.  This might be due to the influence of some blob moving
outward from the  comet nucleus or a shocked region. Such trend is
not seen in December 13 data. Compared to the brightness
distribution on November 6,  it falls sharply close to the nucleus
and then beyond 7" from nucleus, the decrease in the brightness is
gradual. Since we have carried out aperture photometry, it is
difficult to precisely indicate the region of sharp and gradual
fall in the brightness. There is a signature (a kink) in the brightness
distribution  on Dec 13 which might be due to some arc or shocked
region near 10-12" from  the nucleus.

\citet{montalto2008} have studied early motion (ie October 26,
2007 to November 20,2007 period) of the outburst material with
the CCD images in BVRI bands which shows the second blob moving
out from the nucleus. The November 5, 2007 image of 17P/Holmes
\citep{montalto2008} shows  a bright spot close to the
nucleus (cf. their Figure 1) which might have
faded later on. This might explain the rise in  brightness seen
in radial distribution on November 6 in our observations. 

\section{Conclusions}
This work reports on the linear polarisation and the brightness 
distribution observations in the coma of comet 17P/Holmes at phase
angle near 13-14$^\circ$, leading to the following conclusions:

\begin{itemize}

\item Like other comets, 17P/Holmes shows negative polarisation
at  low phase angle (ie below $20^\circ$). Based on the
polarisation values, we believe that 17P/Holmes is not unusual comet as
claimed by \citet{rosenbush2009} and the dust seems to be of the same
nature as found in other comets. The discrepancy could
possibly be due to the gas emission contamination in the broad band
observations.\\

\item Though there is indication that the value of 
polarisation decreases towards longer wavelength, 
we infer that, within the errors, the wavelength dependence of 
polarisation as observed through different sized apertures is typical
of comets.\\

\item  Radial distribution of brightness in the coma shows higher
variation with time in the red wave band as compared to that in
the blue wave band; it has decreased by a factor of 3.6 and 2.5
respectively in red and blue bands indicating increase in the
relative abundance of smaller dust particles away from the
nucleus.\\

\item Radial brightness distribution shows the non-uniform
distribution of material near the comet nucleus on November 6 which
has gradually smoothened by December 13. Also the brightness
distribution has become steeper from November 5-7 run to December 13
run.  
    
\end{itemize}

\section{Acknowledgments}
 We thank the referee for constructive comments and suggestions
that helped to improve the quality of this contribution. We also
thank Christian Buil, Castanet Observatory, France, for allowing us to
use the spectrum of 17P/Holmes, taken by him. The work reported
here is supported by the Department of Space, Government of India. 
\bibliographystyle{mn2e}

\bsp
\label{lastpage}

\end{document}